\documentclass{article}
\usepackage[dvips]{graphicx}
\usepackage{spconf,amssymb,graphicx}
\usepackage{color}
\usepackage{graphicx}
\usepackage{amsmath}
\usepackage{amsthm}
\usepackage{amsfonts}
\usepackage{enumitem}
\usepackage[12pt]{moresize}

\theoremstyle{noparens}

\def\defeq{\stackrel{\triangle}{=}}
\graphicspath{{./figures/} }
\DeclareGraphicsExtensions{.pdf,.jpeg,.png}

\title{Multichannel blind speech source separation with a disjoint constraint source model}
\name{Jianyu Wang$^{1}$, Shanzheng Guan$^{1}$
\address{$^1$CIAIC, Northwestern Polytechnical University, Xi'an, Shaanxi 710072, China \\
}}
\begin{document}
\ninept
\maketitle
\begin{abstract} \vskip 6pt
Multichannel convolutive blind speech source separation refers to the problem of separating different speech sources from the observed multichannel mixtures without much {\em a priori} information about the mixing system. Multichannel nonnegative matrix factorization (MNMF) has been proven to be one of the most powerful separation frameworks and the representative algorithms such as MNMF and the independent low-rank matrix analysis (ILRMA) have demonstrated great performance. However, the sparseness properties of speech source signals are not fully taken into account in such a framework. It is well known that speech signals are sparse in nature, which is considered in this work to improve the separation performance. Specifically, we utilize the Bingham and Laplace distributions to formulate a disjoint constraint regularizer, which is subsequently incorporated into both MNMF and ILRMA. We then derive majorization-minimization rules for updating parameters related to the source model, resulting in the development of two enhanced algorithms: s-MNMF and s-ILRMA. Comprehensive simulations are conducted, and the results unequivocally demonstrate the efficacy of our proposed methodologies.
\end{abstract}

\begin{sloppy}

\section{Introduction}
\label{sec:intro}

Multichannel blind source separation (MBSS) is an unsupervised learning framework to achieve source separation based on a hierarchical generative model of the time-frequency spectrograms of the mixed signals, which can be categorized into underdetermined, determined, and overdetermined cases according to the number of microphones and sources \cite{ozerov2009multichannel, duong2010under, sawada2013multichannel}. The former two refer to the cases where the number of microphones is greater than or equal to the number of sources while the latter refers to the situation in which the number of microphones is smaller than the number of sources.

Independent component analysis (ICA) \cite{comon1994independent} and its extended vector version, i.e., the independent vector analysis (IVA) \cite{kim2006independent}, are the most commonly used approaches in the underdetermined and determined cases. Although these two approaches are widely used in MBSS, they do not consider the spectral structures of sources, which are proven useful for improving  the source separation performance~\cite{kitamura2016determined}. To exploit the source spectral structure, the nonnegative matrix factorization (NMF) \cite{lee2000algorithms} was introduced to MBSS, leading to the so-called multichannel NMF (MNMF). In \cite{ozerov2009multichannel}, NMF was combined with the source model where the covariance matrix
is modeled as a rank-1 matrix. While it is a sound model, this rank-1 assumption makes the resulting algorithm sensitive to reverberation. To deal with BSS in highly reverberant environments, the model with full-rank spatial covariance matrix was proposed \cite{takeda2011underdetermined}. Then, a multiplicative update algorithm was developed to estimate the parameters of source and spatial models \cite{sawada2013multichannel}.

The MNMF methods are computationally expensive. To reduce complexity, the so-called fast full-rank spatial covariance analysis (FastFCA) \cite{ito2021joint} and fast multichannel nonnegative matrix factorization (FastMNMF) \cite{sekiguchi2020fast} were developed based on the assumption that the spatial covariance matrices are diagonal.
For determined MBSS, ILRMA \cite{kitamura2016determined} adopts a rank-1 spatial model to further reduce the computational complexity. It was developed with not only statistical independent assumption between sources but also a low-rank structure of the source spectrograms. This low-rank assumption can help significantly improve the separation performance.
To avoid parameter initialization sensitivity, the so-called $t$-ILRMA \cite{mogami2017independent}, GGD-ILRMA \cite{kitamura2018generalized, ikeshita2018independent, mogami2018independent, mogami2019independent}, and t-MNMF \cite{kitamura2016student} were developed, which generalize the distribution of source model.
To further improve the separation performance, probabilistic sparse distribution was introduced in
\cite{mitsui2017blind,wang2021minimum} to model the dictionary and activation matrices of source spectrograms though this model is still insufficient for accurately modeling the sources in MBSS \cite{takeda2011underdetermined}. Several methods were then developed to leverage the sparseness of the sources in time-frequency domain \cite{o2005survey, gao2010single, feng2018underdetermined,rickard2002approximate}, which have demonstrated some potential to enhance separation performance.

%{\color{red}
%To efficiently address the aforementioned challenges and introduce more precise prior information regarding the source signals, %many methods \cite{o2005survey, gao2010single, feng2018underdetermined} have sought to leverage the sparsity assumption of %sources in the time-frequency domain, specifically, W-disjoint orthogonality of sources \cite{rickard2002approximate}.
%The sparsity assumption aligns with the distribution characteristics of signals in the frequency domain, where only a limited %number of instances exhibit significant values. In general, the energy of a speech signal tends to be concentrated around the %fundamental frequency and its harmonics. Therefore, investigating sparsity priors related to source model parameters holds the %potential to significantly enhance separation performance.
This work is also concerned with how to model the sparesness of the sources, thereby improve the source model in MBSS to further improve source separation performance. We propose to use the Bingham distribution \cite{bingham1974antipodally} for the basis matrices and the Laplace distribution for the activation matrices. This combination forms a hyperspheric-structured sparse regularizer \cite{nadisic2020sparse, abdolali2021simplex, leplat2021multiplicative} applied to MNMF and ILRMA. This regularizer fulfills two purposes: preventing parameter optimization from converging into local optima, and mitigating the issue related to singular values during the optimization process. Simulations are carried out and the results validate the efficacy of the proposed method.

\section{Signal model and problem formulation}
Suppose that there are $N$ sources in the sound field and a microphone array with $M$ sensors  are used to pick up the sound signals. In the short-time Fourier transform (STFT) domain, the multichannel observation signals can be approximately written as
\begin{align}\label{MixSys}
\begin{split}
 \mathbf{x}_{ft} \approx \sum_{n=1}^N \mathbf{a}_{fn} s_{ftn},
\end{split}
\end{align}
where $f$ and $t$ denote, respectively, the frequency and time-frame indices, $m = 1,\dots,M$ and $n = 1,\dots, N$ denote, respectively, the microphone and source indices,
$s_{ftn}$ denotes the signal of the $n$th source at the frequency $f$ and time-frame $t$,
$\mathbf{a}_{fn} \defeq \left[\begin{array}{cccc}a_{fn1} & a_{fn2} & \dots & a_{fnM} \end{array} \right]^T \in \mathbb{C}^{M}$ is the steering vector associated with the $n$th source, and
$\mathbf{x}_{ft} \defeq \left[\begin{array}{cccc} x_{ft1}& x_{ft2} & \dots & x_{ftM} \end{array} \right]^T \in \mathbb{C}^M$  is the multichannel observation signal vector. Note that the additive noise term is neglected in (\ref{MixSys}), so the problem can be specifically formulated for source separation, which has been widely adopted in the literature of BSS.

Let us assume that the signal vector $\mathbf{x}_{ft}$ follows a multivariate complex Gaussian distribution, i.e.,
\begin{align}\label{MCG}
\begin{split}
 \mathbf{x}_{ft} \!\! \sim \!\! \mathcal{N}_{\mathbb{C}}\left(\mathbf{0},{\mathbf{R}}^{(\mathbf{x})}_{ft}\right) \! = \! \frac{1}{\det \left[{\mathbf{R}}^{(\mathbf{x})}_{ft}\right]}\exp\!{\left[\!\! - \! \mathbf{x}_{ft}^H{\left({\mathbf{R}}^{(\mathbf{x})}_{ft}\right)^{\!-1}}\!\!\!\mathbf{x}_{ft} \! \right]}\!,\!
\end{split}
\end{align}
where $\mathbf{R}^{(\mathbf{x})}_{ft} \defeq E \left( \mathbf{x}_{ft}\mathbf{x}_{ft}^H\right)$ is the covariance matrix of $\mathbf{x}_{ft}$, with $E(\cdot)$ denoting  mathematical expectation. The matrix $\mathbf{R}^{(\mathbf{x})}_{ft}$ is Hermitian and is assumed to be positive definite.

The objective of MBSS is to incorporate the information of spatial and source models to estimate the source signal, i.e.,
\begin{align}\label{Demixing}
\begin{split}
% \mathbf{y}_{ft} = \sum_{m=1}^M \mathbf{d}_{fm} {x}_{ftm} = \mathbf{D}_f \mathbf{x}_{ft},
 \mathbf{y}_{ft} = \mathbf{D}_f \mathbf{x}_{ft},
\end{split}
\end{align}
where $\mathbf{y}_{ft} \defeq \left[ \begin{array}{cccc} y_{ft1}&y_{ft2}&\dots&y_{ftN} \end{array} \right]^T \in \mathbb{C}^{N}$ denotes the estimate of the source signals, and
$\mathbf{D}_f = \mathbf{A}_f^{-1} \in \mathbb{C}^{N \times M}$ denotes the demixing matrix.

Let $\mathbf{R}^{(\mathbf{s})}_{fn}$ denote the spatial covariance matrix (SCM). Then, the Sawada's MNMF model given in \cite{sawada2013multichannel} can be expressed as
\begin{align}
\label{SawadaMNMF}
%\begin{split}
 \mathbf{R}^{(\mathbf{x})}_{ft} &=   \sum_{k} \!\! \left( \! \sum_n \mathbf{R}^{(\mathbf{s})}_{fn} z_{nk} \! \right) \!\! w_{fk} h_{kt} \\
 & = \sum_{n} \! \mathbf{R}^{(\mathbf{s})}_{fn} \lambda_{nft}, \nonumber
%\end{split}
\end{align}
where $k = 1,\ldots,K$ is the index of the basis matrices, and $z_{nk}$ denotes the weight for which the $k$th basis belongs to the $n$th source, $w_{fk}$ and $h_{kt}$ denote, respectively, the basis spectra and temporal activations, and $\lambda_{nft} \defeq \displaystyle \sum_{k} w_{fk} h_{kt}$ represents source model parameters.

Now, if it is approximated as a rank-1 matrix, $\mathbf{R}^{(\mathbf{s})}_{ft}$ can be written as the outer product of $\mathbf{a}_{fn}$, i.e.,
\begin{align}\label{rank1SM}
\begin{split}
 \mathbf{R}^{(\mathbf{s})}_{fn} = \mathbf{a}_{fn}\mathbf{a}_{fn}^H.
\end{split}
\end{align}
In this case, the Sawada's MNMF model degenerates to the model used in ILRMA \cite{kitamura2016determined}, in which the covariance matrix $\mathbf{R}^{(\mathbf{x})}_{ft}$ is expressed as
\begin{align}
\label{KitamuraIILRMA}
\begin{split}
 {\mathbf{R}}^{\mathbf{(x)}}_{ft} = &  \sum_{n} \! \mathbf{R}^{(\mathbf{s})}_{fn} \lambda_{nft}
  = \! \sum_n \! \left[ \mathbf{a}_{fn}\mathbf{a}_{fn}^H \! \left( \! \sum_k z_{nk}w_{fk}h_{kt} \! \right) \! \right] \\ = & \mathbf{A}_f \boldsymbol{\Lambda}_{ft} \mathbf{A}_f^H,
\end{split}
\end{align}
where $\boldsymbol{\Lambda}_{ft} \in \mathbb{R}_{\geq 0}^{N\times N}$ is a diagonal matrix with the $n$th diagonal element being $\lambda_{nft} \triangleq \displaystyle \sum_k z_{nk} w_{fk} h_{kt}$.

\section{Proposed Sparseness Model and Source Separation Algorithm}
\subsection{Cost Function}

It is widely known that the priori information of the source model plays an important role in improving MBSS performance \cite{kamo2020regularized, yatabe2021determined, wang2021minimum}. While they have been widely studied, the MNMF model given in
(\ref{SawadaMNMF}) and the ILRMA model given in (\ref{KitamuraIILRMA}) did not consider the sparse structure of the source prior information in the source model. In BSS for speech, the W-disjoint assumption generally holds true, which assumes that in a given time-frequency bin, if one source is dominant, most of energy in the basis coefficient at that bin belongs to that source. This motivates us to propose a source model based on a hyperspheric structure for the sparse priori information, in which the prior over basis matrix is constructed as a Bingham distribution $\mathbf{W}_{nk} \sim \mathcal{B}(\rho_{nk}, \boldsymbol{\theta})$ \cite{bingham1974antipodally}, i.e.,
% over the field of nonnegative real numbers:
\begin{align}\label{HsMBSS}
\begin{split}
 p(\mathbf{W}_{n}\big|\boldsymbol{\rho}_{n}, \boldsymbol{\theta}) = \prod_{k=1}^{K} \frac{1}{C(\boldsymbol{\theta})}  \exp\left( \sum_{f=1}^F \theta_f w_{nfk}^2 - \rho_{nk} \right),
\end{split}
\end{align}
where $\boldsymbol{\rho}_{n} \defeq \left[ \begin{array}{cccc} \rho_{n1}&\rho_{n2}&\dots&\rho_{nK} \end{array} \right]^T \in \mathbb{R}^{K}_{\geq 0}$,  $\boldsymbol{\theta} \defeq \left[ \begin{array}{cccc} \theta_{1}&\theta_{2}&\dots&\theta_{F} \end{array} \right]^T  \in \mathbb{R}^{F}_{\geq 0} $, and $C(\boldsymbol{\theta})$ denotes the parameters with respect to $\boldsymbol{\theta}$. To constrain every basis vector to the unit sphere thereby avoiding singular values, we set the hyperparameter $\boldsymbol{\theta} = \mathbf{1}_F$.

Speech signals in the STFT domain consists of many samples that are far away from the mean. This fact makes Laplace distribution well suited to model speech signals. So, in this work, we use the Laplace distribution $\mathbf{H}_{nk} \sim \mbox{Laplace}(0, \mu_{nk}^{-1})$, i.e.,
%{\color{red} Since the heavy-tail property of the laplace distribution assigns relatively higher probability to extreme values %that are farther from the mean. It is well suitable to model sparsity for speech signals. To further enhance the sparsity of %activation matrix, we use the laplace distribution $\mathbf{H}_{nk} \sim \mbox{Laplace}(0, \mu_{nk}^{-1})$, i.e.,}
\begin{align}\label{HsMBSS2}
\begin{split}
 p(\mathbf{H}_n\big|\boldsymbol{\mu}_n) = \prod_{k=1}^{K} \frac{1}{2} \mu_{nk} \exp \left( - \mu_{nk} \| \mathbf{H}_{nk} \|_1 \right),
\end{split}
\end{align}
where $\boldsymbol{\mu}_{n} \defeq \left[ \begin{array}{cccc} \mu_{n1}&\mu_{n2}&\dots&\mu_{nK} \end{array} \right]^T \in \mathbb{R}^{K}_{\geq 0}$, and $\| \mathbf{H}_{nk} \|_1 = \sum_t| h_{nkt} |$ to enhance the sparsity of the activation matrix in MBSS so that each time-frequency bin in the spectrogram is primarily associated with a few basis elements (columns of basis matrix).

Under the maximum {\em a posteriori} (MAP) framework, one can derive the cost function $Q_{\mbox{s}}$ for the observed multichannel mixed signal $\mathbf{X}$, i.e.,
\begin{align}\label{sMNMF}
%\begin{split}
  &Q_{\mbox{s-MNMF}} = \nonumber \\
  &~~~~ - \log \left[ p(\mathbf{X}\big|\mathbf{R}_{fn}^{(\mathbf{s})}, \mathbf{W}_n,\mathbf{H}_n) p(\mathbf{W}_n\big| \boldsymbol{\rho}_n,\mathbf{1}_F) p(\mathbf{H}_n \big| \boldsymbol{\mu}_n) \right] \nonumber \\
  &~~~~ = \! \sum_{f,t} \! \Bigg[ \! \mbox{Tr} \Big( {\mathbf{R}}^{\mathbf{(x)}}_{ft} (\sum_{n} \mathbf{R}^{(\mathbf{s})}_{fn} \lambda_{nft})^{-1} \! \Big) \! \! + \! \log \det (\sum_{n} \mathbf{R}^{(\mathbf{s})}_{fn} \lambda_{nft}) \Bigg] \!  \nonumber \\
  &~~~~ + \! \sum_{n,k} \mu_{nk} \| \mathbf{H}_{nk} \|_1 + \sum_{n} \left( \mathbf{1}_F^T\mathbf{W}_{n}^{\cdot 2} - \boldsymbol{\rho}_n^T\right) + \mbox{Cst},
%\end{split}
\end{align}
where $\mbox{Cst}$ is a const, $Q_{\mbox{s}}$ denotes the cost function for MNMF with the hyperspheric structure for the sparse priori information (s-MNMF).

If the rank-1 approximation is used, (\ref{sMNMF}) then degenerates to the hyperspheric structure based ILRMA model (s-ILRMA), i.e.,
\begin{align}
\label{sILRMA}
  &Q_{\mbox{s-ILRMA}}  =  \! \sum_{f,t} \! \Bigg[ \! \mbox{Tr} \Big( \! {\mathbf{y}_{ft}^H} \mathbf{D}_f^{-H} \! \big( \mathbf{D}_{f}^H \boldsymbol{\Lambda}_{ft}^{-1} \mathbf{D}_{f} \! \big) \mathbf{D}_{f}^{-1} {\mathbf{y}_{ft}} \! \Big) \! \nonumber \\
  &~~~~ \! - \! T\sum_{f} \log|\mathbf{D}_{f}\mathbf{D}_f^H| \! + \! \sum_{f} \sum_{t} \log|\boldsymbol{\Lambda}_{ft}|  \nonumber \\
  &~~~~ + \sum_{n,k} \mu_{nk} \| \mathbf{H}_{nk} \|_1 + \sum_{n} \left(\mathbf{1}_F^T\mathbf{W}_{n}^{\cdot 2} - \boldsymbol{\rho}_n^T\right) +\mbox{Cst}.
\end{align}

\subsection{Source Separation Algorithm}
The logarithmic determinant and the trace terms in \eqref{sMNMF} and \eqref{sILRMA} make it difficult to optimize the two cost functions, i.e., $Q_{\mbox{s-MNMF}}$ and $Q_{\mbox{s-ILRMA}}$, directly. To develop the optimization algorithm, we first discuss the upper bound for the two objective functions. Once the minimum of the upper bound is determined under certain rule, the cost function is nonincreasing under the same rule. According to \cite{sawada2012efficient, yoshii2016student, leplat2020blind, sekiguchi2019semi}, we have the following two inequalities.
\begin{itemize}[leftmargin=4mm]
 \itemsep=0pt
  \item For the concave function $f(\mathbf{V}) \! = \! \log\det(\mathbf{V}) \! \quad \! (\mathbf{V} \! \in \! \mathbb{C}^{N\times N})$, it satisfies:
      \begin{align}\label{TaylorExp}
        \begin{split}
            \! f(\mathbf{V}) \! = \! \log\det\mathbf{V} \leq   \! \log\det\widehat{\mathbf{V}} \! + \! \mathrm{Tr}(\widehat{\mathbf{V}}^{-1}\mathbf{V}) \! - \! N,
        \end{split}
      \end{align}
      where $\widehat{\mathbf{V}}$ is an arbitrary positive semi-definite matrix, and the equality holds if $\mathbf{V} = \widehat{\mathbf{V}}$.

        \item For the convex function $g(\mathbf{V}) = \mathrm{Tr}(\mathbf{V}^{-1} \widehat{\mathbf{V}})$, it satisfies
      \begin{align}\label{ConcaveIneq}
        \begin{split}
            g(\{\mathbf{V}_n\}_{n=1}^{N}) = & \mathrm{Tr}\Bigg( \bigg( \sum_{n=1}^N \mathbf{V}_n \bigg)^{-1} \mathbf{K} \Bigg) \\
            & \leq \sum_{n=1}^N \mathrm{Tr}(\mathbf{V}_n^{-1}\boldsymbol{\Phi}_n \mathbf{K} \boldsymbol{\Phi}_n^H),
        \end{split}
      \end{align}
      where $\mathbf{K}$ is any positive semi-definite matrix, $\{\mathbf{V}_n\}_{n=1}^{N}$ is a set of matrices, and $\{\boldsymbol{\Phi}_n\}_{n=1}^{N}$ is a set of auxiliary matrices.
\end{itemize}

The above two inequalities have been used to form auxiliary functions in BSS \cite{fevotte2009nonnegative, fevotte2011algorithms}. In this work, we also adopt the two inequalities to relax the two functions $Q_{\mbox{s-MNMF}}$ and $Q_{\mbox{s-ILRMA}}$. Substituting the two inequalities in \eqref{TaylorExp} and \eqref{ConcaveIneq} into \eqref{sMNMF}, we can derive the upper bound for $s$-MNMF in \eqref{sMNMF}, which is denoted as $\mathcal{U}_{\mbox{s-MNMF}}$, i.e.,
\begin{align}
\label{UppersMNMF}
& \mathcal{U}_{\mbox{s-MNMF}} \! = \! \sum_{n,f,t} \!\! \left[ \lambda_{nft}^{-1} \mbox{Tr} \bigg( \!\! \left(\mathbf{R}_{fn}^{(s)}\right)^{-1} \!\! \boldsymbol{\Phi}_{ftn} \mathbf{R}_{ft}^{(\mathbf{x})} \boldsymbol{\Phi}_{ftn}^H  \bigg) \right]  \nonumber \\
&~~~~ + \sum_{n,f,t} \lambda_{nft} \mbox{Tr} \left( \mathbf{R}_{fn}^{(s)} \left( \widehat{\mathbf{R}}_{ft}^{(\mathbf{x})} \right)^{-1} \right) + \sum_{f,t} \log\det{\widehat{\mathbf{R}}_{ft}^{(\mathbf{x})}} \nonumber \\
&~~~~  +  \sum_{n,t} \boldsymbol{\mu}_{n}^T\mathbf{H}_{nt}  + \sum_{n} \left(\mathbf{1}_F^T\mathbf{W}_{n}^{\cdot 2} - \boldsymbol{\rho}_n^T\right),
\end{align}
where $\mathbf{H}_{nt}$ denotes the column vector of $\mathbf{H}_{n}$, $\boldsymbol{\Phi}_{ftn}$, and $\widehat{\mathbf{R}}_{ft}^{(\mathbf{x})}$ denote the auxiliary variables. When $\boldsymbol{\Phi}_{ftn}$ and $\widehat{\mathbf{R}}_{ft}^{(\mathbf{x})}$ satisfy $\boldsymbol{\Phi}_{ftn} = \left( \lambda_{nft} \mathbf{R}_{fn}^{(s)} \right) \left( \sum_{n} \lambda_{nft} \mathbf{R}_{fn}^{(s)} \right)^{-1}$ and $\widehat{\mathbf{R}}_{ft}^{(\mathbf{x})} = \sum_n \lambda_{nft} \mathbf{R}^{(\mathbf{s})}_{fn}$, respectively, the above upper bound is tight.

Identifying the partial derivative of \eqref{UppersMNMF} with respect to $\mathbf{H}_{n}$ and forcing the result equal to zero gives
\begin{align}\label{PartialUppersMNMFH}
\begin{split}
 & - \sum_f h_{nkt}^{-2} w_{nfk}^{-1} \mbox{Tr} \bigg( \left( \mathbf{R}_{f,n}^{(s)} \right)^{-1} \boldsymbol{\Phi}_{ftn} \mathbf{R}_{ft}^{(\mathbf{x})} \boldsymbol{\Phi}_{ftn}^H  \bigg) \\
 & + \sum_{f} w_{nfk} \mbox{Tr} \left( \mathbf{R}_{fn}^{(s)}  \left( \widehat{\mathbf{R}}_{ft}^{(\mathbf{x})} \right)^{-1} \right) + \mu_{nk} = 0.
\end{split}
\end{align}
It follows immediately that we can have the following iterative estimate for $h_{nkt}$:
\begin{align}\label{sMNMFupdateH}
\begin{split}
\vspace{-5cm}
 h_{nkt} \! \leftarrow \! h_{nkt}^\prime \!\!\! \sqrt{\frac{ \! \sum_f \! w_{nfk} \!\! \mbox{Tr} \bigg( \!\! \mathbf{R}_{fn}^{(s)} \!\! \left( \!\! \widehat{\mathbf{R}}_{ft}^{(\mathbf{x})} \!\! \right)^{\!\!-1} \!\!\! \mathbf{R}_{ft}^{(\mathbf{x})} \!\! \left( \!\! \widehat{\mathbf{R}}_{ft}^{(\mathbf{x})} \!\! \right)^{\!\!-1}  \!\! \bigg)}{\sum_{f} w_{nfk} \mbox{Tr} \left( \mathbf{R}_{fn}^{(s)} \left( \!\! \widehat{\mathbf{R}}_{ft}^{(\mathbf{x})} \!\! \right)^{\!\!-1} \right) + \mu_{nk}}} ,
\end{split}
\end{align}
where $h^\prime_{nkt}$ denotes the estimate of $h_{nkt}$ in the previous iteration step.

Similarly, identifying the partial derivative of \eqref{UppersMNMF} with respect to $\mathbf{W}_n$ and forcing the result equal to zero gives% $\frac{\partial \mathcal{U}_{\mbox{s-MNMF}}}{\partial {w}_{nfk}} = {0}$. Then, we obtain:
\begin{align}
\label{PartialUppersMNMFW}
 & - \sum_t w_{nfk}^{-2} h_{nkt}^{-1} \mbox{Tr} \bigg( \left( \mathbf{R}_{fn}^{(s)} \right)^{-1} \boldsymbol{\Phi}_{ftn} \mathbf{R}_{ft}^{(\mathbf{x})} \boldsymbol{\Phi}_{ftn}^H  \bigg) \nonumber \\
 & + \sum_{t} h_{nkt} \mbox{Tr} \left( \mathbf{R}_{fn}^{(s)} \left( \!\! \widehat{\mathbf{R}}_{ft}^{(\mathbf{x})} \!\! \right)^{\!\!-1} \right) + 2 w_{nfk} = 0.
\end{align}
According to \cite{rechtschaffen200892}, \eqref{PartialUppersMNMFW} can be rearranged into the form of a linear equation with one variable, which can then be solved by the cubic-root method.

The update rule for the parameters of the spatial model $\mathbf{R}_{fn}^{(s)}$ can also be obtained through the similar process in \cite{sekiguchi2019semi}  [equations from (57) to (60)], i.e., identifying the partial derivative of \eqref{UppersMNMF} with respect to $\mathbf{R}_{fn}^{(s)}$, and forcing the result equal to zero \cite{yoshii2018correlated}.

Similarly, one can derive the following auxiliary function for $s$-ILRMA \cite{fevotte2011algorithms,kitamura2016determined}:
%, we derive the  \cite{fevotte2011algorithms} of \eqref{sILRMA} according to \cite{kitamura2016determined}:
\begin{align}\label{UppersIILRMA}
& \mathcal{U}_{\mbox{s-ILRMA}} = \sum_{n,t} \boldsymbol{\mu}_{n}^T\mathbf{H}_{nt}  +  \sum_{n} \left(\mathbf{1}_F^T\mathbf{W}_{n}^{\cdot 2} - \boldsymbol{\rho}_n^T\right) \nonumber \\
&~~~~ + \sum_{n,f,t,k} \frac{|{y_{nft}}|^2 \alpha_{nftk}^2}{w_{nfk}h_{nkt}} - T\sum_{f} \log|\mathbf{D}_{f}\mathbf{D}_f^H|  \nonumber \\
&~~~~ + \sum_{n,f,t} \frac{1}{\beta_{nft}} \bigg( \sum_k w_{nfk} h_{nkt} - \beta_{nft} \bigg) + \log{\beta_{nft}},
\end{align}
where $\boldsymbol{\mu}_n \defeq \left[\begin{array}{cccc} \mu_{n1}& \mu_{n2} & \dots & \mu_{nK} \end{array} \right]^T \in \mathbb{R}^K$, $\alpha_{nftk} \geq 0$ is auxiliary variables, which satisfy $\sum_k \alpha_{nftk} = 1$,  and $\beta_{nft} \geq 0$ are also auxiliary variables. If $\alpha_{nftk} = \frac{w_{nfk}h_{nkt}}{\lambda_{nft}}$ and $\beta_{nft} = \lambda_{nft}$, the upper bound of $s$-ILRMA degenerates to to \eqref{sILRMA}.

It follows that the parameters of the source model $w_{nfk}$ and $h_{nkt}$ in $s$-ILRMA can be updated as follows:
\begin{align}\label{Update_SourceH}
\begin{split}
 h_{nkt} \leftarrow h_{nkt}^\prime \sqrt{\frac{\sum_{f} w_{nfk} |y_{nft}|^2 \lambda_{nft}^{-2} }{\sum_{f} w_{nfk} \lambda_{nft}^{-1} + \mu_{nk} }}.
\end{split}
\end{align}
The parameter $w_{nfk}$ can also be obtain by solving a cubic equation of one variable and the result is
\begin{align}\label{Update_SourceW}
\begin{split}
 w_{nfk} \! \! \leftarrow \! \frac{ \! \sqrt{ \! \left( \! \sum_t \! h_{\!n\!k\!t} \! \right)^2 \! \! \! + \! 8 w_{\!n\!f\!k}^\prime \! \sum_t \! \frac{|y_{nft}|}{\lambda_{nft}} h_{nkt} } \! \! - \! \sum_t \! h_{nkt} \! }{4}.
\end{split}
\end{align}

%For the spatial model,
The demixing matrix $\mathbf{D}_f$ in $s$-ILRMA is updated based on the rules of AuxIVA, which are as follows:
\begin{align}\label{Update_ILRMAD}
 \widehat{\mathbf{R}}^{(\mathbf{s})}_{fn} & = \frac{1}{T}\sum_t\frac{1}{\lambda_{nft}}\mathbf{x}_{ft}\mathbf{x}_{ft}^H, \\
 \mathbf{d}_{fm} & \leftarrow (\mathbf{D}_f\widehat{\mathbf{R}}^{(\mathbf{s})}_{fn})^{-1}\mathbf{e}_m, \\
 \mathbf{d}_{fm} & \leftarrow \mathbf{d}_{fm}(\mathbf{d}_{fm}^H\widehat{\mathbf{R}}^{(\mathbf{s})}_{fn}\mathbf{d}_{fm})^{-\frac{1}{2}},
\end{align}
{where $\widehat{\mathbf{R}}^{(\mathbf{s})}_{fn}$ denotes an auxiliary variable, $\mathbf{d}_{fm}$ is a row vector of $\mathbf{D}_f$, and $\mathbf{e}_m$ denotes the $m$th column vector of an $M\times M$-dimensional identity matrix.}

\section{Simulations}
\subsection{Simulation Setup}
The configuration of the SISEC challenge \cite{araki20122011} is adopted in this work to generate simulation data for evaluating the developed MBSS algorithms with $M = N = 2$. The clean speech signals are taken from the Wall Street Journal (WSJ0) corpus \cite{garofolo1993csr}. The room size is set to $8~\mathrm{m}~\times~8~\mathrm{m}~\times~3~\mathrm{m}$. Two microphones are placed at the center of the room with a spacing of $2.83$ cm and the same hight. Two loudspeakers are also positions the same height as the microphones but are $2$ meters away from the center of the two microphones to simulate two sources. The incident angles of the two sources were randomly selected from $[0^{\circ},90^{\circ}]$ and $[0^{\circ},-90^{\circ}]$ respectively per mixture, where the direction normal to the line connecting the two microphones is $0^{\circ}$. The image source model \cite{allen1979image} is used to generate the room impulse responses, where the sound absorption coefficients were calculated by the Sabine's formula \cite{young1959sabine} with the specified room size and the reverberation time $T_{60}$, which ranges from $0$ to $600$ ms with an interval of $50$ ms. For each gender combinations (there are four combinations) and every value of $T_{60}$, we generated 100 mixtures for evaluation. The sampling rate is $16$ kHz.

All the coefficients of $\boldsymbol{\rho}_n$ in $s$-MNMF and $s$-ILRMA are set to $10$ and all the coefficients of  $\boldsymbol{\mu}_n$ in $s$-MNMF and $s$-ILRMA are set to $0.05$. Besides $s$-MNMF and $s$-ILRMA, the following algorithms are also evaluated for the purpose of comparison: $s$-ILRMA with AuxIVA \cite{ono2011stable}, MNMF \cite{sawada2013multichannel}, $m$-MNMF \cite{wang2021minimum}, ILRMA \cite{kitamura2016determined}, $t$ILRMA \cite{mogami2017independent}, sub-Gaussian distributed ILRMA (sGD-ILRMA) \cite{ikeshita2018independent} and $m$-ILRMA \cite{wang2021minimum}. Signal-to-distortion ratio (SDR) and source-to-interference ratio (SIR) \cite{vincent2006performance} are adopted as the performance metrics.

\begin{figure}[t]
\vspace{-0.3cm}
\begin{minipage}[b]{.46\linewidth}
  \centering
  \centerline{\includegraphics[width=4.2cm]{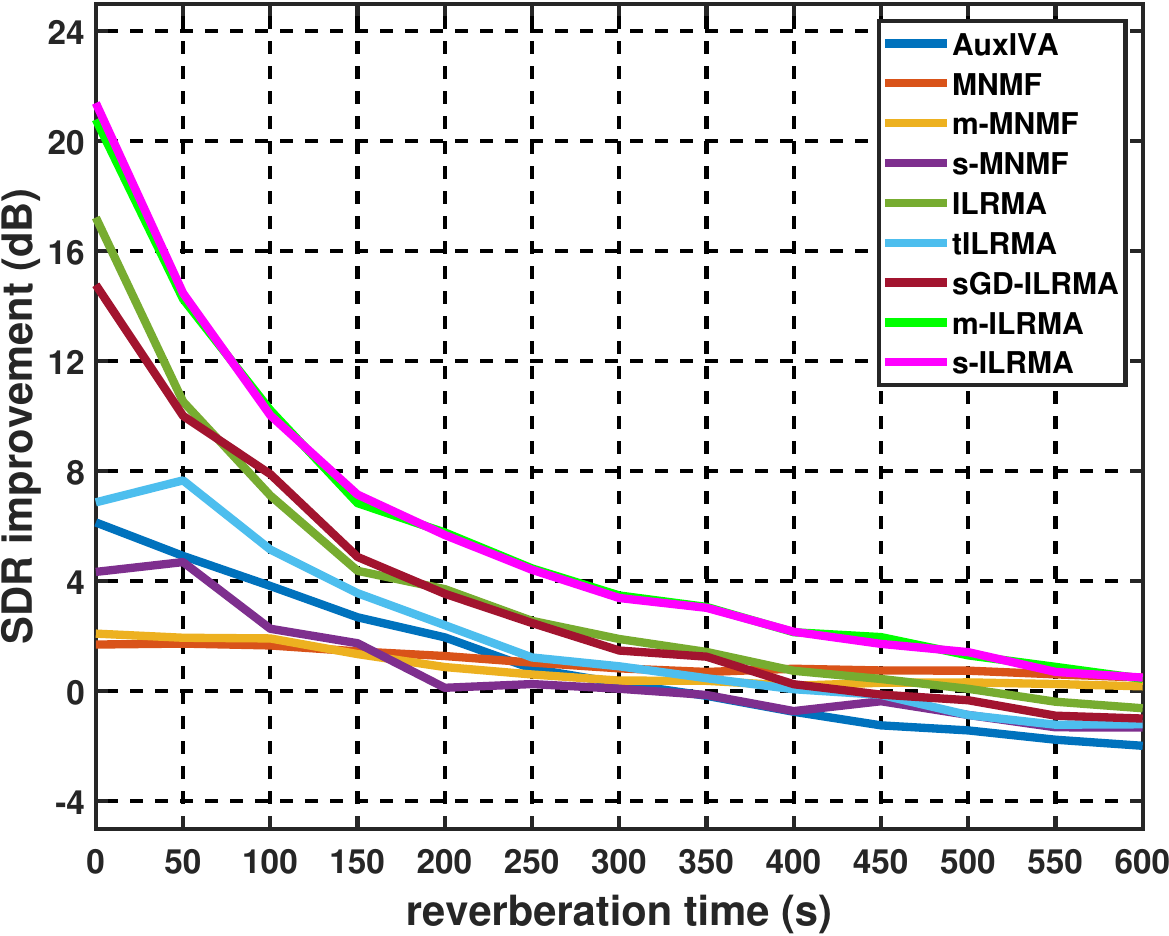}}
%  \vspace{2.0cm}
  \centerline{(a) female+female}\medskip
\end{minipage}
\begin{minipage}[b]{.55\linewidth}
  \centering
  \centerline{\includegraphics[width=4.2cm]{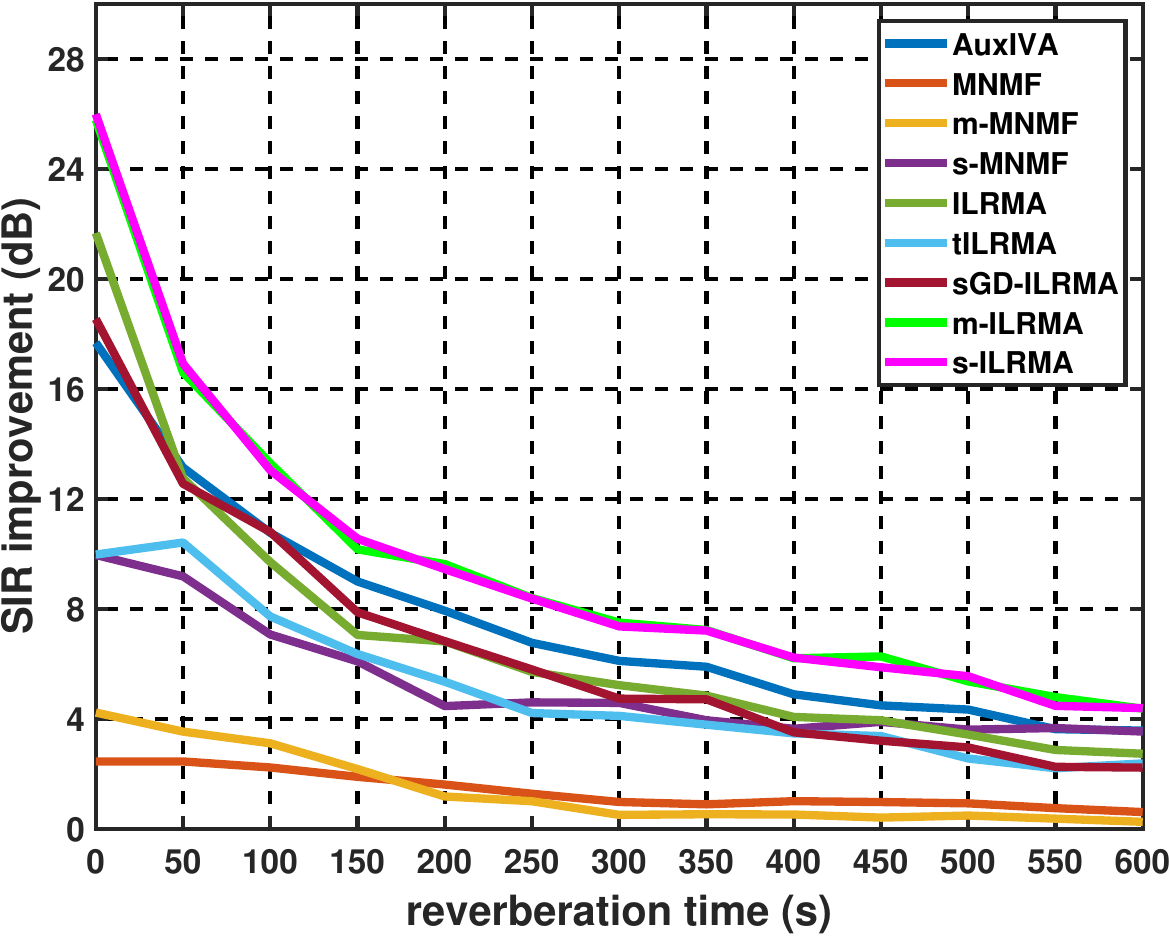}}
%  \vspace{1.5cm}
  \centerline{(b) female+female}\medskip
\end{minipage}
%\hfill
\vspace{-0.5cm}

\begin{minipage}[b]{0.46\linewidth}
  \centering
  \centerline{\includegraphics[width=4.2cm]{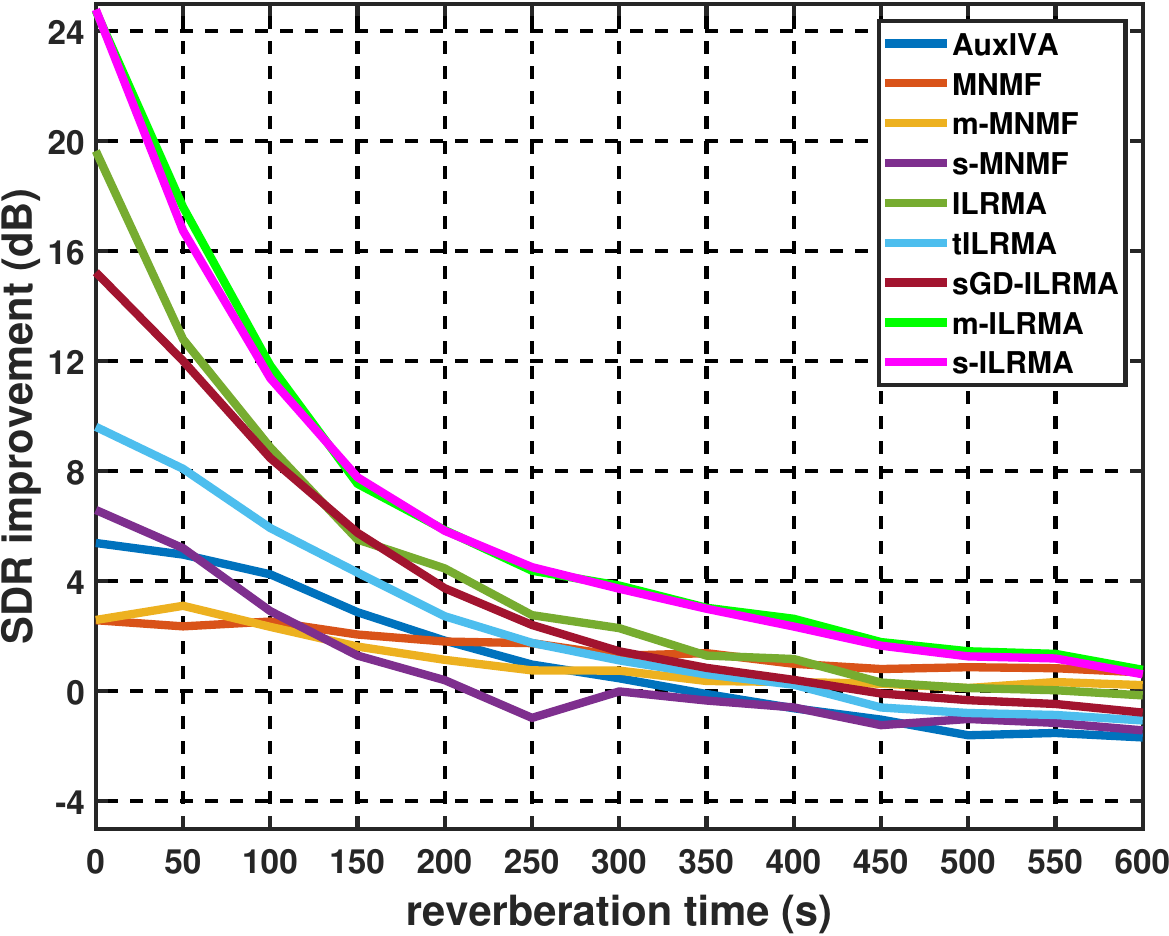}}
%  \vspace{1.5cm}
  \centerline{(c) male+male}\medskip
\end{minipage}
\begin{minipage}[b]{.55\linewidth}
  \centering
  \centerline{\includegraphics[width=4.2cm]{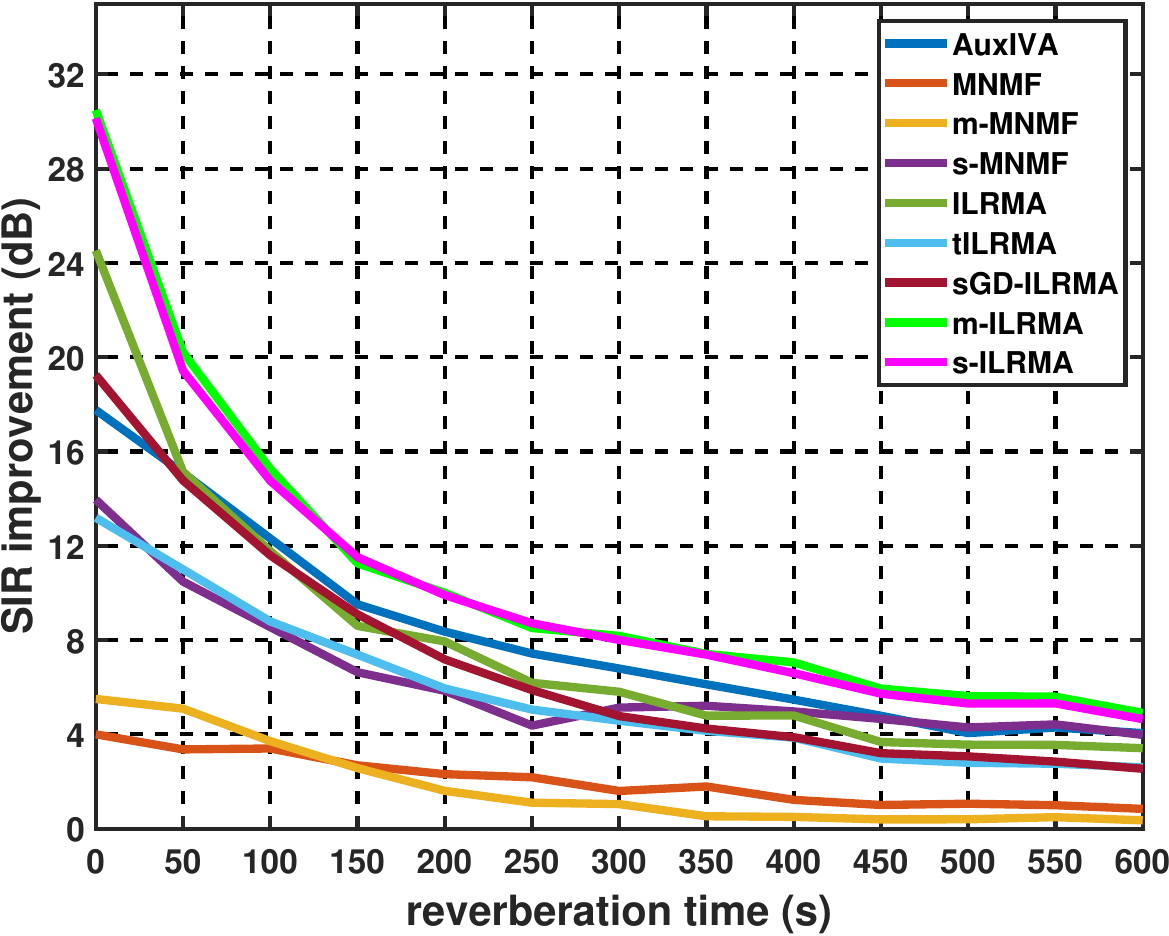}}
%  \vspace{2.0cm}
  \centerline{(d) male+male}\medskip
\end{minipage}
\vspace{-0.5cm}

\begin{minipage}[b]{.46\linewidth}
  \centering
  \centerline{\includegraphics[width=4.2cm]{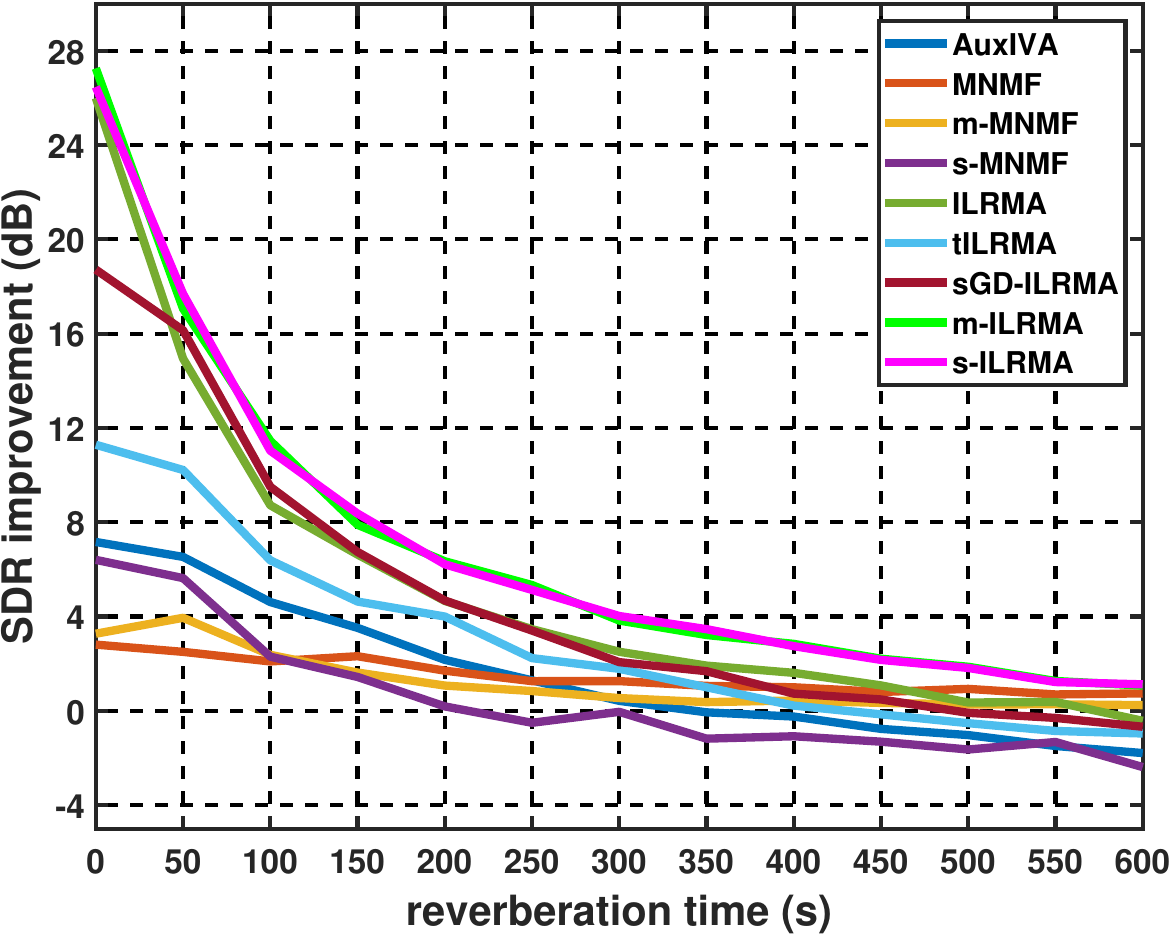}}
%  \vspace{1.5cm}
  \centerline{(e) female+male}\medskip
\end{minipage}
%\hfill
\begin{minipage}[b]{0.55\linewidth}
  \centering
  \centerline{\includegraphics[width=4.2cm]{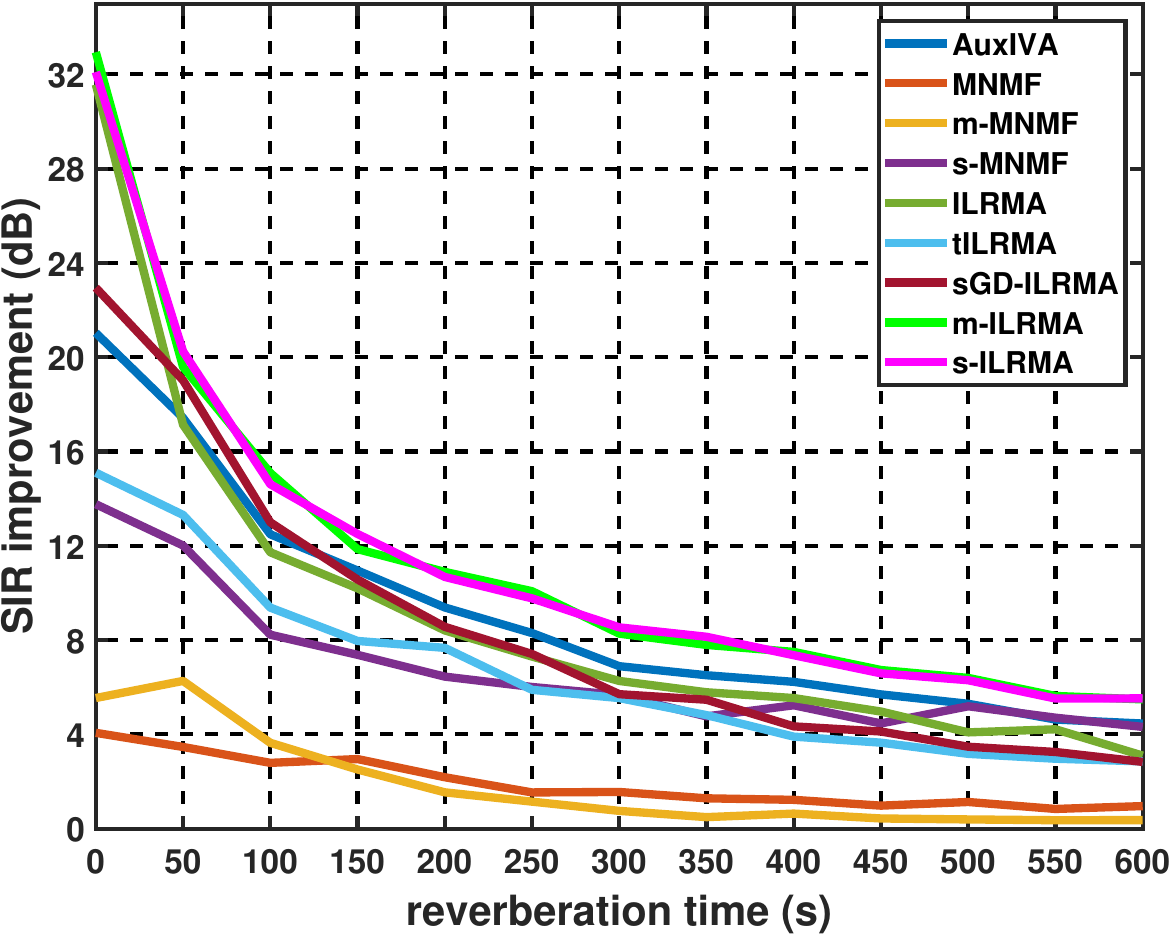}}
%  \vspace{1.5cm}
  \centerline{(f) female+male}\medskip
\end{minipage}
%\vspace{-0.8cm}
\caption{SDR and SIR improvements of the studied methods.}
\label{fig:2}
\vspace{-0.45cm}
\end{figure}

\subsection{Simulation Results}
Figure~\ref{fig:2} plots the SDR and SIR improvement over the mixed speech in different reverberation conditions. One can make the following observations from the results. First, the performances of the $s$-ILRMA and $m$-ILRMA algorithms are similar, which are consistently better than the performance of the other studied methods. On average, the $s$-ILRMA algorithm achieved an additional SDR gain of approximately $1.7$ dB, $5.3$ dB, and $4.9$ dB, respectively, as compared to ILRMA, $m$-MNMF, and AuxIVA. Second, The SDR and SIR improvement yielded by $s$-MNMF is larger than that by MNMF and $m$-MNMF in light reverberant environments (e.g., $T_{60}$$\le$100~ms), in which average SDR improvements of $s$-MNMF are $3.0$ dB and $2.5$ dB higher than MNMF and $m$-MNMF, respectively.

\section{Conclusions}

This papers developed two source separation algorithms: $s$-MNMF and $s$-ILRMA, which are improved versions of MNMF and ILRMA, respectively. These two algorithms constrain, respectively, the MNMF and ILRMA algorithms with the hyperspheric-structured sparse regularizer to improve the sparsity of the source model estimation. To solve the constrained optimization problems,
we relaxed the logarithmic determinant terms with their tightened lower bounds and derived the multiplicative update rules for parameters optimization. Simulation were carried out and the results show that the $s$-MNMF algorithm outperforms MNMF and  $m$-MNMF and the $s$-ILRMA algorithm outperforms AuxIVA, ILRMA, $t$-ILRMA, and sGD-ILRMA, which are four representative blind source separation algorithms well studied in the literature of BSS.

\vfill\pagebreak

%\bibliographystyle{IEEEbib}
%\bibliography{strings,refs}

\bibliographystyle{IEEEbib}
%\bibliography{strings,refs}

\end{sloppy}

\end{document}